\documentclass{sigplan-proc}
\usepackage[latin1]{inputenc}
\usepackage{color,setspace}
\usepackage{amsmath}
\usepackage{fullpage}
\usepackage{graphicx}
\usepackage{float}
\usepackage{pst-all}
\usepackage{moreverb}
\usepackage{verbatim}
\newfloat{algorithm}{thp}{lop}[section]
\floatname{algorithm}{Algorithm}
\newfloat{module}{thp}{lop}[section]
\floatname{module}{Module}

\newcommand{\remove}[1]{}
\newtheorem{definition}{Definition}
\newtheorem{theorem}{Theorem}
\newtheorem{lemma}{Lemma}
\newtheorem{corollary}{Corollary}
\newtheorem{note}{Note}
\newlength {\squarewidth}

\newcounter{linecounter}

\newenvironment{theorem-repeat}[1]{\begin{trivlist}
\item[\hspace{\labelsep}{\bf\noindent Theorem~\ref{#1} }]}%
{\end{trivlist}}

\newenvironment{lemma-repeat}[1]{\begin{trivlist}
\item[\hspace{\labelsep}{\bf\noindent Lemma~\ref{#1} }]}%
{\end{trivlist}}

\begin{document}

\author{Davide Canepa \hspace{1cm} Maria Gradinariu Potop-Butucaru\\
Université Pierre et Marie Curie (Paris 6), CNRS,INRIA, France\\
canepa.davide@tiscali.it, maria.gradinariu@lip6.fr}
\date{ }
\title{Self-stabilizing Tiny Interaction Protocols} 

\maketitle

\begin{abstract}
In this paper we present the self-stabilizing implementation  
of a class of token based algorithms. In the current work we only consider interactions 
between weak nodes. They are uniform, they do not have unique identifiers, are static and 
their interactions are restricted to a subset of nodes called neighbours. 
While interacting, a pair of neighbouring nodes may create mobile 
agents (that materialize in the current work the token abstraction) 
that perform traversals of the network and accelerate the system stabilization. 
In this work we only explore the power of oblivious stateless agents. 
%We prove that they are sufficient to implement a large class of self-stabilizing 
%token based schemes designed for anonymous distributed systems.  
Our work shows that the agent paradigm is an elegant distributed tool for achieving 
self-stabilization in Tiny Interaction Protocols (TIP). Nevertheless, in order to reach 
the full power of classical self-stabilizing algorithms more complex 
classes of agents have to be considered (e.g. agents with memory, identifiers or communication skills).
Interestingly, our work proposes for the first time a model that unifies the recent 
studies in mobile robots(agents) that 
evolve in a discrete space and the already established population protocols paradigm. 
\end{abstract}

%\newpage
\pagenumbering{arabic}
%\doublespace
\section{Introduction}
Recently, the distributed computing community started to investigate the 
interactions in biological and chemical systems in order to provide efficient 
computational models for adhoc systems like sensor or 
peer-to-peer. One of the most promising research in this direction 
is the population protocol model developed by Angluin 
{\it et al.} through a series of papers 
\cite{AngluinADFP2006,AAFJ08}.
In this model, 
pairs of nodes chosen by an adversary interact and change their state
according to a transition function.
In \cite{AngluinADFP2006} it is stated that for each such transition function,
the resulting population protocol is said to stably 
compute a predicate on the initial states of the nodes if,
after sufficiently many interactions in a fair execution, all nodes converge to having the correct value of the
predicate. 
Nodes in this framework have no identity and no ability to distinguish  
two different interactions with the same node.  
Interactions between nodes in this model have various incarnations.
Sometime interactions are restricted by the choice of 
a fair or randomized scheduler while in other situations the network topology is 
the main parameter that defines them. This corresponds perfectly to the real networks.
In sensor networks, for example, the topology (the geographical position of nodes) is the main parameter having 
a major impact on sensor interactions while in peer-to-peer networks the interactions are restricted to the virtual overlay defined by the peers. 

The main concern was to evaluate the computational power of the population protocols model and 
several problems have been addressed: coloration, phase 
synchronization, counting, leader election \cite{AngluinADFP2006,AAE08a,AAFJ08}.
Fault tolerance and security issues in variants of the model have been addressed
in \cite{DFGR06,DFGR07}. Self-stabilizing population protocols have 
been further addressed.
One of the first problems that has been investigated was leader election 
and the first work that addresses this issue \cite{AAFJ08}.
Interestingly, the self-stabilizing extension of classical population 
protocols quickly meet dead-ends. Several impossibility results 
have been proved. Therefore,
\cite{AAFJ08} extends the study to the non-uniform 
leader election in odd and directed rings. Furthermore, due to the persisting impossibility results 
the model is enriched with fairness assumption and oracles 
(abstractions that offer some global information).
%The algorithm foresee probes that the Leaders send in order to kill the other Leaders and 
%assumes global fairness. 
Fisher and Jiang study the self-stabilizing leader-election problem in 
this model in \cite{DBLP:conf/opodis/FischerJ06}. 
They introduce $\Omega?$  an oracle that reports true or false if it 
detects the presence or the absence of a leader. 
Using $\Omega?$, the authors provide uniform 
and self-stabilizing leader election algorithms for fully
 connected networks under the assumption of local fairness and for rings under the global fairness assumption.
In \cite{DBLP:conf/opodis/FischerJ06} the authors also prove that uniform 
 leader election is impossible in rings assuming 
local fairness, even with the help of $\Omega?$. In the current work, we prove that this result holds 
even under global fairness without additional assumptions. 
In \cite{CIW09} the authors investigate the memory necessary to solve the problem without oracles.

Another problem addressed in the classical model of population protocols is the 
 self-stabilizing token circulation. This problem has been addressed only in non-uniform 
population protocols \cite{AAFJ08}. Our work extends the study to the uniform population protocols.
We prove the possibility of deterministic solutions only on chains and 
using global fairness assumptions.
For the general graphs we propose an impossibility result and a probabilistic 
solution.

These studies prove that the self-stabilization in 
population protocols model reached a dead-end when dealing 
with both static (e.g. leader election) or 
dynamic tasks (e.g. token circulation). Therefore, a recent variant of the model explores 
the self-stabilization of the system when the interactions benefit from the presence 
of a base station with incorruptible infinite memory. This model, extremely powerful by 
its hypothesis already proved its effectiveness since most of the 
tasks self-stabilizing in classical distributed settings found also a solution 
in this model \cite{BBK09}. However, the main challenge to 
address would be the minimum assumptions one has 
to make on the interaction system in order to reach its stabilization. 
One natural idea is the use of the popular paradigm of mobile agents. This paradigm already 
proved its efficiency in optimizing the self-stabilization 
in classical distributed settings \cite{ghosh,jb}.

Interestingly, the ``probes'' used in the Fisher's leader election algorithm to ``destroy'' 
the other leaders in the population are very similar
to the agents paradigm. Therefore, the extension of population protocols with 
mobile agents seems a promising research direction. 
%Therefore the extension of the population protocols with agents 
%came as a natural idea.
%The mobile agents paradigm, very popular in the study of ambient intelligence, already 
%prove its effectiveness in self-stabilizing classical distributed algorithms 
%\cite{jb,elad, ghosh}. 
Our goal is to unify the population protocols model and the agents paradigm.
We advocate that the computational power of the 
population protocols can benefit from the agents ability to ``move'' from one node 
to another and hence to disseminate or gather information. 
The model becomes even more interesting when agents are allowed to 
execute some code when they are guested by a particular node.

\subsection{Our contribution}
We propose a novel model of interactions that extends the 
population protocols with the power of agents and oracles schemes.
In the current work we explore the self-stabilization 
power of the weakest version of this model, TIP (Tiny Interaction Protocols): the 
interactions between nodes is restricted to a 
predefined neighbourhood, nodes and agents are oblivious, anonymous and uniform.  
In this model we consider the self-stabilization of 
both dynamic and static tasks. More specifically, we consider two popular case studies 
in self-stabilization: the token circulation and the leader 
election (in its local and global version).

Our contribution  is threefold. First, we propose some negative results 
related to the feasibility of leader election
and token circulation in the TIP model.  
Then, we present some positive results: self-stabilizing 
solutions for token circulation in chains and arbitrary graphs.
Furthermore, we propose solutions for self-stabilizing 
local leader election.
Finally, we prove for both static and dynamic tasks 
the necessity of an oracle that provides to each node 
information about the current state of the system. Interestingly, we prove that 
in the case of the token circulation this information has to be global in both 
deterministic and probabilistic settings 
while for the case of the
local leader election only a local knowledge is sufficient. 
We also show that in the case of the global leader election 
even the global knowledge is not sufficient to solve the problem. 
In this case, additional 
assumptions (e.g. memory on agents or nodes) have to be made.

\subsection{Paper Road-map}
The paper is organized as follows. 
Section \ref{sec:model} proposes de TIP model. 
In Section \ref{sec:eld} we provide some impossibility results related to the 
token circulation in TIC. 
Sections \ref{sec:token} proposes deterministic and probabilistic
solutions for token circulation in some particular classes of graphs. 
In Section \ref{sec:mis} we address the 
local leader election and propose a negative result for 
the global leader election even with the help of the global fairness.

\section{Model}
\label{sec:model}

\subsection{Interaction Protocols with mobile agents}
% We propose an overview of the extension we propose to the population protocol model. 
%Some of the definitions proposed below are borrowed from
%$\cite{AngluinADFP2006, DBLP:conf/podc/AngluinADFP04,DBLP:conf/dcoss/AngluinFJ06}$. 

We represent a network by a fixed undirected graph $G = (V,E)$. Each vertex represents a 
finite-state sensing device and an edge ($u$, $v$) 
 indicates the possibility of interaction  between ${u}$ and ${v}$ in which ${u}$
 is the {\it initiator} and ${v}$ is the {\it responder}. For a node $u$, 
$\mathcal{N}_u$ is the set of all the nodes $v_i$ responder of ${u}$. 

Each node may have a local set of variables and a memory slot reserved for storing one or several agents. An agent is a mobile abstraction (e.g. message, mobile program) that can be locally created, deleted, modified, executed or sent by a node to another node. 
Each node can access a local primitive, LocalAgent?() which returns \emph{true} if locally there is an agent. 
When interacting with another node, a node can create a local agent, destroy the local agent, execute the local agent or push the local agent to the peer node in the interaction.

A local state of a node is given by the value of its local variables 
and the state of the local agent if LocalAgent?() invocation returns true. 
The system can be modelled as a transition system. When two nodes $u$ and $v$ interact, their state changes atomically from  $(s_u,s_v) \rightarrow (s^\prime_u,s^\prime_v)$, referred in the following as \emph{local transition}. A local transition may be either deterministic or probabilistic.  
A configuration of the system at some time $t$ is the set of states of the nodes in the system at $t$. Contrary to existing models for population protocols our model allows several pair of nodes to interact in a given configuration. The only restriction we impose is that concurrent interactions are pairwise independent. That is, two pairs of nodes that interact at the same time $t$ have no common node. 
Let $c_t$ be the configuration of the system at time $t$. There is a \emph{global transition} of the system at time $t$ from $c_t$ to $c_{t+1}$ if there is at least one local transition in $c_t$. 
An execution of the system, $e$, is a sequence of configurations $e=(c_1, c_2, \ldots)$ where $c_{i+1}$ is obtained from $c_i$ by a global transition. 

\subsection{Schedulers and Fairness}
Intuitively, a scheduler in population protocols chooses the pairs 
of agents that will interact in a given configuration. 
Formally a scheduler is a predicate over the executions of the system.
In this paper we consider the weakly fair version of the following
schedulers : arbitrary and k-bounded. 
A scheduler is weakly fair if, in an infinite execution, a continuously enabled
pair of agents is eventually activated.    
\begin{itemize} 
\item \emph{$k$-bounded}: between two consecutive activations of a
pair of agents, another pair of nodes can be activated at most $k$~times;
\item \emph{arbitrary}: at each configuration an arbitrary subset of
  pair of nodes is activated.
\end{itemize}

Note that contrary to some work done in population protocols 
the above schedulers are not randomized. A randomized
scheduler is one of the weakest schedulers one may consider
therefore this type of schedulers are not the object of
the current work. 

We also consider a stronger version of the above defined weak 
fairness referred in 
the population protocols literature as \emph{global fairness}. 
With global fairness an interaction that is infinitely often possible 
in an execution is infinity often scheduled.

\subsection{Faults and Self-stabilization}
In this paper we assume that nodes can start their execution in any
configuration.  For the particular case of 
token circulation or leader election the faulty period may cause the
lost or the creation of the agent that materializes the token and nodes have no
possibility to detect locally this faulty state. 
In order to deal with this kind of 
faults we use oracles and self-stabilization
tools.

 A self-stabilizing system \cite{Dol00} started in an arbitrary configuration eventually
 exhibits a correct behaviour according to its specification. 
%We advocate
%that self-stabilization is an appealing framework for dealing with
%faults in sensor-networks where the local memory of sensors may
%be easily corrupted/perturbed by external devices.

\begin{definition}[self-stabilizations]
Let $\mathcal{L_{A}}$ be a non-empty
\emph{legitimacy predicate} (legitimacy predicate is 
defined over the configurations of a system and 
is an indicator of its correct behaviour) of an algorithm $\mathcal{A}$ with respect to a
specification predicate $Spec$ such that every configuration satisfying
$\mathcal{L_{A}}$ satisfies $Spec$.
Module $\mathcal{A}$ is \emph{self-stabilizing} with respect to $Spec$ iff
the following two conditions hold:\\
\textsf{(i)} Every computation of $\mathcal{A}$ starting from a configuration
satisfying $\mathcal{L_A}$ preserves $\mathcal{L_A}$ (\emph{closure}).  \\
\textsf{(ii)} Every computation of $\mathcal{A}$ starting from an arbitrary configuration
contains a configuration that satisfies $\mathcal{L_A}$
(\emph{convergence}).
\end{definition}

The merge between population protocols and failure detectors was made
for the first time in \cite{DBLP:conf/opodis/FischerJ06} where 
an eventual leader oracle (eventual leader detector), $\Omega?$ is 
introduced to solve leader election. This oracle is useful when the
system is started in symmetric configurations (no leader is elected). 
Note that the eventual leader detector $\Omega?$ is a weaker version of the oracle $\Omega$ introduced 
first in \cite{CHT96} and proved to be the weakest failure detector to solve consensus. Instead 
of electing a leader (as $\Omega$ does), $\Omega?$ reports to each node
whether or not at least one leader is present in the network. Note that the guess may be correct 
or not and different guesses may be reported to different nodes. The only guarantee offered is 
that from some point onward if there is continuously a leader or if there is continuously no leader,
$\Omega?$ eventually accurately reports this fact to all nodes.

In this paper, we will use the eventual agent detector. Similar to the
eventual leader detector defined in
\cite{DBLP:conf/opodis/FischerJ06}, 
the agent detector reports if at least one agent is present in the
network.

\begin{definition}[Eventual agent detector]
The {\it eventual agent detector}, $A?$ supplies a Boolean input to each node
at each step so that the following conditions are satisfied by every execution $e$:
\begin{itemize}
\item If all but finitely many configurations of $e$ lack of agent, then each node receives 
input {\it false} at all but finitely many steps.
\item If all but finitely many configurations of $e$ contain one or more agents, then each node receives 
input {\it true} at all but finitely many steps.
\end{itemize}
\end{definition}
 
In mobile robots, \cite{FIPS10}, a similar abstraction is used: the unlimited 
robots visibility.
The originality of our approach is to address the geographical power of the oracle.
Interestingly, for some tasks it is sufficient that 
the oracle provides only a local information 
(e.g. for the case of the local leader election it is sufficient only 
the one hop distance information as  discussed later in the paper). 
In the case of the global leader 
election we prove that this problem is impossible to solve even if 
this oracle offers information on the whole network.

\subsection{Leader election and Unique Token circulation}
In this paper we address two well known problems in distributed
computing : leader election and token circulation. 
These two problems are similar in the sense that they share the safety property
: a unique token/leader should be present 
in the system in any configuration 
(a token is a predicate over the local configurations of a node). However, the liveness part is
different.  In the leader election the unique token should be hold by the 
same node and no other node in the system should hold the token in 
the subsequent configurations while in the token circulation the unique token has to perpetually 
visit every node in the system.
In our study, the token will be materialized by an agent.

\begin{definition}[unique token legitimate configuration]
A  configuration is legitimate for the unique token iff exactly one
node holds the token in this configuration. 
\end{definition}

\begin{definition}[Unique Token Circulation]
 A system is self-stabilizing for the token circulation specification iff
(i) each execution of the system converges in a finite number of steps 
to a unique token legitimate configuration and (ii) each process in the
system holds the unique token infinitely often.
\end{definition}

\begin{definition}[Silent Leader Election]
A system is silently self-stabilizing for the leader election specification iff
each execution of the system converges in a finite number of steps 
to a unique token legitimate configuration and no node is enabled
in that configuration.
\end{definition}

In this paper we also address the local leader election which 
restricts the election to a neighbourhood. That is, 
each process has to have a unique 
leader in its neighbourhood. Note that the local leader election is a weaker version of the 
MIS problem which focus on optimizing the set of local elected leaders to the minimal set.  
Recent local algorithms that addressed this problem are proposed in \cite{Lenzen2009Local}. 
Interestingly the transformation of these 
algorithms in the population protocols model has not been explored yet.
 
%\paragraph{Eventual Leader Oracle}

\subsection{Work hypothesis}
In the current paper we assume a network of small devices with a static topology. 
Nodes  do not have unique identification. The interactions between nodes follow 
the interactions model described above and are restricted 
by the topology of the network and the scheduler choice. 
During their interactions nodes can create agents that may further change their locations.
A node invokes the LocalAgent?() primitive in order to detect the 
local presence of the agent. In the following we assume the weakest 
class of agents: anonymous and memoryless. That is, the agents do not carry 
any memory or code to be executed by their hosts.   
The only operations nodes can execute during an interaction: check if they hold locally an agent, 
create an agent or delete/push the local agent to the peer node. 
We also assume that each node receives Boolean inputs 
from the eventual agent oracle that reports true if at least one agent is present 
in the network or false otherwise.   
This system is referred in the following \emph{TIP (Tiny Interaction Protocols)}.

\section{Token Circulation in TIP}
\subsection{Impossibility results for token circulation in TIP}
\label{sec:eld}
In the following we show the necessity of additional assumptions in
order to provide uniform solutions for self-stabilizing
leader election or token circulation in TIP (Tiny Interaction Protocols). Notice 
that memory is an important factor that may help 
bypassing many of the impossibility results stated below however
additional memory means additional corruptions so the system should
pay additional time and effort in order to be stabilized.
The following note restates in the context of the new interaction 
model results already known in the classical distributed systems.

\begin{note}
%\textit{{\bfseries Lemma 4.} 
\label{lemma:4}
Let $\mathcal S$ be a TIP. 
It is impossible to guarantee the presence of
a unique agent in $\mathcal S$ without
additional assumptions.
%The leader detector $\Omega?$ is necessary to implement leader
%election in $\mathcal S$.
\end{note}

\begin{figure}[!]
\begin{center}
 \begin{picture}(240,40)
   \put(10,15){\circle{20}}
   \put(5,30){\makebox(10,10){A}}
   \put(60,15){\vector(-1,0){39}}
   \put(65,30){\makebox(10,10){B}}
   \put(70,15){\circle{20}} 
   \put(80,15){\vector(1,0){39}} 
   \put(125,30){\makebox(10,10){C}}
   \put(130,15){\circle{20}} 
   \put(140,15){\vector(1,0){39}}
   \put(185,30){\makebox(10,10){D}}
   \put(190,15){\circle{20}}  
   \end{picture}
   
   \begin{picture}(240,40)
   \put(10,15){\circle{20}}
   \put(5,30){\makebox(10,10){A}}
   \put(60,15){\vector(-1,0){39}}
   \put(65,30){\makebox(10,10){B}}
   \put(70,15){\circle{20}} 
   \put(80,15){\vector(1,0){39}} 
   \put(125,30){\makebox(10,10){C}}
   \put(130,15){\circle{20}} 
   \put(140,15){\vector(1,0){39}}
   \put(185,30){\makebox(10,10){D}}
   \put(190,15){\circle*{35}}  
   \end{picture}
\end{center}
\caption{Two initial configurations for Note \ref{lemma:4}}
\label{fig:lemma4}
\end{figure}
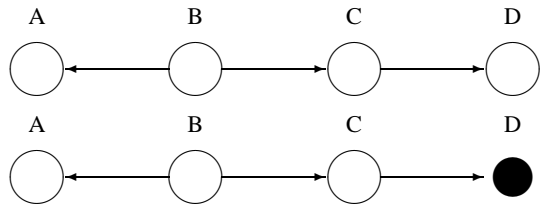

%\begin{proof}
%\textit{Proof.} 
%$\Omega?$ plays a key role in the creation of new leaders. 
The intuition of the above result is as follows: 
without additional assumptions it is impossible to 
decide if the system already has a unique agent, 
hence new agents may be introduced infinitely and system 
never converges to a  configuration with a unique agent.
That is, consider a chain topology and two initial configurations 
one without any agent, the other with an agent hold by D (see Figure
\ref{fig:lemma4}). Since the system can start in any configuration
both configurations may be initial configurations 
for a legal execution of the system.

Consider node B. It can interact only with nodes A and C. 
Nodes A an C hold no agent in both configurations so in the B's view
these two configurations are identical. The following cases can occur:

\begin{itemize} 
\item B introduces a new agent in the first configuration. Since B has the
same view in both configurations, B will execute the same action in
the second configuration as well. This will 
transform the second configuration in an illegitimate one since  
it will contain two agents.   
\item B does not introduce a new agent and no other node becomes agent holder. The first configuration is
illegitimate and stays illegitimate for ever.  
\end{itemize}
Hence, we can exhibit infinite executions that never 
converge to a unique mark legitimate configuration. 
%Overall, without $\Omega?$ it's impossible 
%to decide if or not a new leader should 
%be introduced in the system.
%\end{proof}   

\begin{note}
Note \ref{lemma:4} does not hold for a system with two nodes. 
In this case a simple self-stabilizing 
algorithm is the following: if neither the initiator nor the responder
are agent holders then one of them create an agent; if both the initiator 
and the responder are agent holders
then one of them becomes agent free. 
\end{note}

In the following we prove that in TIP with 
general acyclic graph topology, 
self-stabilizing unique token
circulation is impossible even with the help of 
$A?$ and the global fairness assumption. 
%Note that the result also holds for silent leader election.

%{\bf Note that for chains even if the unicity of the token is
%guaranteed, tokens should travel in the networks forever so the
%silent part is not guaranteed}.

\begin{figure}
\centering
\includegraphics[scale=.4]{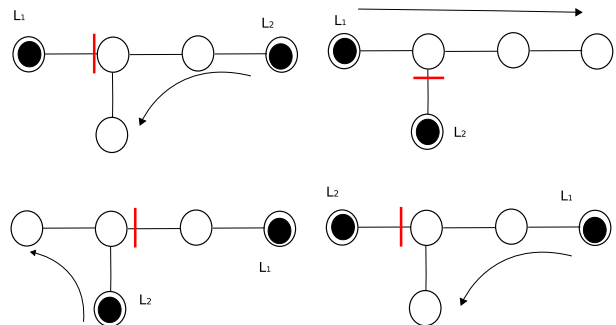}
\caption{Scheduling for the impossibility result of Lemma \ref{lemma:impag}}\label{fig:1}

\end{figure}

\begin{lemma}
\label{lemma:impag} 
Let $\mathcal S$ be a TIP with arbitrary topology. No deterministic self-stabilizing 
unique agent circulation algorithm exists in 
$\mathcal S$, even with the help of agent detector $A?$ 
and the global fairness assumption.  
\end{lemma}

\begin{proof}
Assume a self-stabilizing deterministic unique agent circulation algorithm for general acyclic graphs, that works under global 
fairness with the help of $A?$.
Consider a  graph with two agents $L_1$ and $L_2$ (see Figure \ref{fig:1}).
Call  the nodes with more than two edges \textit{traffic lights}. These \textit{traffic lights} are always 
red in the direction of one of the two agents, so one agent cannot enter 
the  \textit{traffic light} node while the other one can cross all the edges of that node but the red one. 
In such conditions we show that deterministic unique agent circulation is impossible, even 
under the global fairness assumption. 
Due to the red light the two agents never interact. 
Thanks to the $red / light$, the graph is divided in two parts. 
By the fairness assumption each agent visits each node of its component. 
Since the fair scheduler changes the direction of the $red / light$ infinitely many times, 
each agent visits each node infinitely often  without ever interact with the other agent Figure \ref{fig:1}. 
Moreover the agent detector $A?$ becomes useless because the $red  \ light$ works regardless of its indications.
Since the two agents never interact the unique agent circulation behaviour is never verified.
\end{proof}

%\begin{lemma}
%\label{interactions}
%Let $\mathcal S$ be a system of weak agents and acyclic topology.  
%There is no deterministic self-stabilizing leader election.
%\end{lemma}

%\begin{proof}
%\end{proof}

%In the following we define the {\it sibling leader detector} which behaves as $\Omega?$ 
%for nodes that are not in the leader state. Nodes that are in the leader state receive $true$
%if there is another leader in the system or $false$ otherwise.
%A possible implementation of the sibling leader detector is as follows. 
%A leader detects the presence of another leader, 
%if they interact directly (i.e. they are initiator and/or responder in a interaction), 
%or indirectly, if one of them interacts with probes of the other leader. 

%Lemma \ref{lemma:impossibility}  and \ref{interactions} 
%show that necessarily in any leader election algorithm leaders should perceive 
%the presence of other leaders.

\section{Self-stabilizing Unique Token Circulation}
\label{sec:token}
In this section we propose deterministic and probabilistic 
solutions for token circulation in population protocols with weak
agents. 

%\paragraph{Data Structures.}
Each node can hold either an agent that will represent the token abstraction
$\spadesuit$ or nothing $-$ (following the result of the 
LocalAgent?() invocation) and each node receives its current input true (T) or false (F) 
from $A$?. $A$? returns (T) when at least one agent is
present in the network and (F) when no agent is present.

\subsection{Deterministic Unique Token Circulation on Chains}
In the following we consider chain topologies and propose self-stabilizing 
deterministic algorithms for token circulation under global fairness.

Intuitively the algorithm works as follows. 
A clean node(a node without token) 
becomes agent holder, when the agent detector 
signals the absence of any agent in the system (Rule 2).
When two nodes holding an agent each interact, the responder becomes clean (Rule 1). 
If the responder has an agent and the initiator is a clean node, the latter creates an 
agent and the former becomes clean (Rule 3). Otherwise, no state change occurs. 
%Each node outputs L when it holds a $\spadesuit$ , otherwise it outputs N.
Note that the wild-card symbol, $*$, is used to replace any value.

\begin{algorithm}[h]
%\textbf {Algorithm 1}

\begin{verse}
\textit	{	\textbf	{Rule 1. } 	$((\spadesuit,*),( \spadesuit,*)) \longrightarrow  ((\spadesuit),(-)) $	}\\ 
\textit	{	\textbf	{Rule 2. } 	$((-,F),(-,*)) \longrightarrow  ((\spadesuit),(-)) $   }\\
\textit	{	\textbf	{Rule 3. } 	$((-,*),(\spadesuit,*)) \longrightarrow  ((\spadesuit),(-))$ 	}
\end{verse}
\caption{Unique Token Circulation on chains}
\label{alg:letrees}
\end{algorithm}

%\begin{definition}[legitimate configuration]
%{\bfseries Definition 1.}
%\label{legitimatechain} 
%A configuration of Algorithm \ref{alg:letrees} is legitimate for a
%chain topology if and only if only there is only one token in the network.
%\end{definition}

\begin{lemma}
\label{lemma:convergence}
Let ${\mathcal S}$ be a TIP system. Algorithm \ref{alg:letrees} converges 
to a legitimate configuration for unique token circulation under asynchronous 
scheduler and global fairness assumptions.
\end{lemma}

\begin{proof}
Let $e$ be an execution of Algorithm \ref{alg:letrees} starting in 
an illegitimate configuration, $c$.
The following situations are possible.
\begin{itemize}
\item There is no agent in $c$. In this case, 
all pairs of nodes are enabled for
Rule 2 and the scheduler has to chose at least one pair of these
nodes. After their execution at least one agent is
introduced in the system. Due to the fairness assumption the agent will visit each node of the network.
\item There are several agents in $c$. For the sake of simplicity we assume two agents. In a chain 
topology, Rule 3 and the fairness assumption make each agent visit all nodes.  
Assume the two agents never meet. This is
equivalent to say that there is at least a node that is never visited 
by an agent which is impossible by the global fairness 
assumption. When the two tokens become neighbours the execution of
Rule 1 reduces the number of agents becomes 1 and the proof reduces to the first case.     
\end{itemize}
\end{proof}

\begin{lemma}
Let ${\mathcal S}$ be a TIP system executing Algorithm \ref{alg:letrees}.
${\mathcal S}$ self-stabilizes to the token circulation 
specification under an asynchronous scheduler and global
fairness assumptions.
\end{lemma}

\begin{proof}
Following Lemma \ref{lemma:convergence}, 
$S$ converges to a legitimate
configuration in a finite number of steps. 
By the fairness assumption the unique agent in the network will
visit each node infinitely often.
\end{proof}

\subsection{Self-stabilizing Token Circulation in Arbitrary Graphs}
\label{sec:leag}
In the following we propose a probabilistic self-stabilizing algorithm
that solves the unique token circulation using the agent detector $A?$. The algorithm 
works under k-bounded scheduler.
The algorithm idea is as follows. Agents perform random walks in order
to find and destroy other agents.  If no agent is reported by
$A?$ then new agents are introduced in the system.

%\subsection{Detailed description}
%Each node has a memory slot that can hold a bit with two states: the token mark $ \spadesuit$ or $-$. 
%Each node receives as input the current value reported by $A?$.
A clean node creates an agent when there is no agent in the system $A?$=F (Rule 2). 
If two agent holders interact (one of them as initiator and the other
as responder), the responder looses its agent (Rule 1). 
If an agent holder interacts with a clean node, the agent is moved from the initiator to the responder 
with a probability of 1/2 (Rule 3) and if the initiator is
clean and the responder has an agent, then the agent moves 
with probability 1/2 from the latter to the former. Rule 2 introduces agents 
when $A?$ reports their absence. Rule 1 destroys extra 
agents. Rule 3 allows agents to travel in the network in order 
to meet each other and eventually be destroyed via Rule 1.

\begin{algorithm}[h]
%   \textbf{Algorithm 2}
   
   \begin{quote}In the following we discuss the leader election feasibility in TIP.
We increase the power of the system by adding an eventual
%$\star$-mark 
agent detector.

	\textit	{	\textbf	{Rule 1. } 	$((\spadesuit,*),( \spadesuit,*)) \longrightarrow  ((\spadesuit),(-)) $	}\\
    \textit	{	\textbf	{Rule 2. } 	$((-,F),(-,*)) \longrightarrow  ((\spadesuit),(-)) $   } \\
         $$\textit	{	\textbf	{Rule 3. } 	$((\spadesuit,*),(-,*)) \longrightarrow $}
	\left\lbrace
				\begin{array}{cc}  Pr(1/2) & ((-),(\spadesuit))\\	
				   				    Pr(1/2) & ((\spadesuit),(-))
				\end{array}
	\right.$$
    %    $$\textit	{	\textbf	{Rule 4. } 	$((-,*),(\spadesuit,*)) \longrightarrow$} 
	%\left\lbrace
	%			\begin{array}{cc}  Pr(1/2) & ((-),(\spadesuit))\\	
	%			   				    Pr(1/2) & ((\spadesuit),(-))
	%			\end{array}
	%\right.$$

	\end{quote}
\label{alg:leproba}
\caption{Probabilistic agent circulation}					
\end{algorithm}

%\begin{definition}[Legitimate configuration]
%The system executing Algorithm \ref{alg:leproba} is in a legitimate 
%configuration if it verifies the unique token circulation legitimate
%predicate and $A?$ outputs true to every node in the system.
%\end{definition}

\begin{lemma} 
\label{lemma:9} An agent covers infinitely often a virtual ring that includes all nodes in the system.
\end{lemma}   

\begin{proof} 
Assume there is a node of the graph that is never visited by the agent.
Either the agent is blocked in a node or the agent cycles in a part of the graph.
In the first case the agent holder is enabled for the Rule 3.
The probability for this node to keep the agent infinitely 
is 0: $lim_{s\rightarrow\infty} [(\frac{1}{2})^s]$.
In the second case, either 
the agent is pushed back and forth between two nodes or the agent
travels in a cycle. Both cases are impossible due to the fairness assumption.   
\end{proof}

\begin{corollary}
Two agents that cover two virtual rings visit at least one common
node.
\end{corollary}

\begin{lemma} 
\label{lemma:10} Let e be an execution of Algorithm \ref{alg:leproba} starting in a 
configuration with two agents. Eventually, two agent holders interact 
under the k-bounded scheduler assumptions.
\end{lemma}

Due to space limitation the proof of the above lemma is not provided.
The proof uses similar arguments as the correctness of 
classical self-stabilizing token based schemes \cite{BGJ07}.
%\begin{proof} 
%Let $p$ be a node of the system and let $A_1$ and $A_2$ two agents.
%Lemma \ref{lemma:9} proves that eventually the token mark $A_1$ visits $p$ and the token 
%mark $A_2$ also visits $p$.
%Let $d_1$ be the smallest distance between $A_1$ and $p$ and let $d_2$ be the smallest 
%distance between $A_2$ and $p$. Due to fairness assumptions eventually the distance between 
%$T_1$ and $p$ decreases until $T_1$ reaches $p$. $T_2$ eventually reaches a neighbor of $p$ 
%in a finite number of steps. Meantime, $T_1$ does not change its position (we assume a bounded scheduler).
%Once $T_1$ and $T_2$ are neighbors, by Rule 1, one of the two tokens marks is destroyed. 
%\end{proof}

 \begin{lemma} 
Let $e$ be an execution of Algorithm \ref{alg:leproba}
 starting in an arbitrary configuration. 
$e$ converges to a legitimate configuration.
\end{lemma}

\begin{proof} 
Suppose there are no tokens in the initial configuration of $e$. So
from some point on, every node receives false from 
$A?$. By Rule 2 the initiators declare themselves agent holders 
and the system reaches a configuration with one ore more agents.
Starting from this configuration, some clean nodes may receive false from 
their detector and continue to inject agents but there is a point in the execution from which 
$A?$ returns true to every node in the system.
From this point onward no new agents are injected in the system.
Suppose the system in a configuration with more than two agents and $A?$ returns true to every node in the system. 
Let $k$ be the number of agents in this configuration.
By Lemma  \ref{lemma:10} two agents in this set eventually interact and by Rule 1 one of them disappears.
So starting from a configuration with $k$ agents in a finite number of steps the number of agents drops to $k-1$. 
The process is iterated until the system reaches a legitimate configuration.
\end{proof}

The correctness of our system is a direct consequence of the previous lemmas.
\begin{theorem}
The system executing Algorithm \ref{alg:leproba} verifies the token
circulation specification assuming a bounded scheduler.
\end{theorem}

\section{Global and Local Leader Election in TIP}
\label{sec:mis}
In the following we discuss the leader election feasibility in TIP.
We increase the power of the system by adding an eventual
%$\star$-mark 
agent detector. We address both the global leader election and its local version.

\subsection{Impossibility results related to leader election in TIP}
\begin{lemma} 
\label{lemma:impossibilitygle}
Let $\mathcal S$ be a TIP.
There is no deterministic or probabilistic uniform self-stabilizing
silent leader-election algorithm in $\mathcal S$ even with the help of 
an eventual agent detector $A?$ without additional assumptions.
\end{lemma}

 \begin{proof} 
Intuitively the proof goes as follows.
Suppose the presence of two leaders (each leader holds an agent) and none of them can notice the
  existence of the other one. Even with the help of an agent detector 
$A$?, for each of them it is impossible  
to decide if it is the only leader or there is another leader in the system.
  
Consider two configurations, $c$ and $c_1$, and a chain topology: one 
with two leaders in nodes B and D and the other one 
with a leader in node B (see Figure
\ref{fig:lemma-impossibility}). Note that $c_1$ is a terminal
configuration since it is legitimate.
 
\begin{figure}[!]
\begin{center}
 \begin{picture}(240,40)
   \put(10,15){\circle{20}}
   \put(5,30){\makebox(10,10){A}}
   \put(60,15){\vector(-1,0){39}}
   \put(65,30){\makebox(10,10){B}}
   \put(70,15){\circle*{50}} 
   \put(80,15){\vector(1,0){39}} 
   \put(125,30){\makebox(10,10){C}}
   \put(130,15){\circle{20}} 
   \put(180,15){\vector(-1,0){39}}
   \put(185,30){\makebox(10,10){D}}
   \put(190,15){\circle*{50}}  
   \end{picture}
   
   \begin{picture}(240,40)
   \put(10,15){\circle{20}}
   \put(5,30){\makebox(10,10){A}}
   \put(60,15){\vector(-1,0){39}}
   \put(65,30){\makebox(10,10){B}}
   \put(70,15){\circle*{50}} 
   \put(80,15){\vector(1,0){39}} 
   \put(125,30){\makebox(10,10){C}}
   \put(130,15){\circle{20}} 
   \put(180,15){\vector(-1,0){39}}
   \put(185,30){\makebox(10,10){D}}
   \put(190,15){\circle{20}}  
   \end{picture}
\end{center}
\caption{Two initial configurations for Lemma \ref{lemma:impossibilitygle}}
\label{fig:lemma-impossibility}
\end{figure}
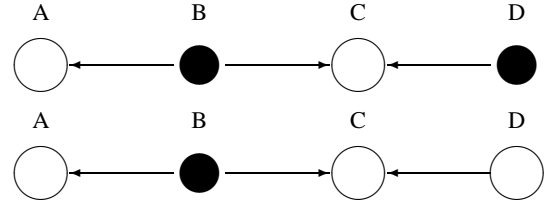   
 Node B has visibility only on its neighbors and $A?$ can notify only if there is at least an agent 
 in the network, so from its point of view, the two configurations are
identical.
The following cases arise: 
\begin{itemize}
\item B holds its agents. Since B has the same view in both
configurations, the first configuration is still illegitimate
since it has two leaders.
\item B becomes agent free. Since B has the same view in
both configuration, the same action is executed in both
configurations. Two new configurations are obtained:
$c^\prime$ and $c_1^\prime$ and in $c_1^\prime$ there is no leader. 
If in $c_1^\prime$, B and D decide to become leader, since to both of them $\Omega?$ 
returns false, the system returns to a configuration similar to the
initial configuration and the system is not any more silent.
\item B pushes the leader mark to one of its neighbors (say C) and C
may do the same since it has the same ``view'' as $B$ in the previous 
configuration. So, the leader mark arrives on $D$ which has the same
view in both the configurations (the legitimate and illegitimate configuration).
\end{itemize}
Overall, even helped by $\Omega?$ it's impossible to assert 
if the leader-election configuration is reached or not without
additional assumptions.
\end{proof}

\begin{lemma} 
\label{lemma:impossibilitylle}
Let $\mathcal S$ be a TIP with ring topology of odd size.
There is no deterministic or probabilistic uniform self-stabilizing
silent local leader-election algorithm in $\mathcal S$ 
without additional assumptions.
\end{lemma}

\begin{proof}
Consider a ring topology of odd size and the following initial configuration $(n,n,a,n,a,\ldots,a)$ where $n$ denotes an empty node and $a$ denotes an agent holder. Whatever the fair scheduling the configuration cannot stabilize 
to a configuration where clean node alternate with agent nodes since the size of the ring is odd.
\end{proof}

In order to bypass the impossibility result for the case of odd size 
rings we add a k-distance agent detector, $A^k?$. Differently from the global 
agent detector the k-distance agent detector reports if up to distance
$k$ there is an agent. Algorithm \ref{alg:lle-odd-rings} implements
local leader election in general graphs
%odd sized rings with 
using $A^1?$. The
algorithm idea is the same as for the even sized rings. The main
difference comes in the interaction of clean nodes. They introduce a
new agent only and only if they have no agent in their neighbourhood.
Each node execution the algorithm has either an agent agent $\spadesuit$ (the LocalAgent?() returns true) 
or is empty and
receives the input of $A^1?$.

\begin{algorithm}[h]
%\textbf {Algorithm 1}

\begin{verse}
\textit	{	\textbf	{Rule 1. } 	$((\spadesuit,*),( \spadesuit,*)) \longrightarrow  ((\spadesuit,*),(-,*)) $	}\\ 
\textit	{	\textbf	{Rule 2. } 	$((-,F),(-,*)) \longrightarrow  ((\spadesuit,*),(-,*)) $   }\\
\end{verse}
\caption{Local leader election for even sized rings}
\label{alg:lle-odd-rings}
\end{algorithm}

\begin{lemma} Algorithm \ref{alg:lle-odd-rings} is a silent implementation
of local leader election in general graphs under asynchronous scheduler and 
global fairness assumption.
\end{lemma}

\begin{proof} 
Intuitively, the proof goes as follows.
Let $T$ be the set of conflicting pair of neighbors. That is, 
either both have an agent or they are clear and have no neighbors with an agent.
Due to the fairness assumption, each of these pair of nodes will eventually interact and apply 
either Rule 1 or Rule 2. After each 
interaction the size of $T$ eventually decreases. In a finite number of interactions the systems stabilizes.
\end{proof}

\section{Conclusions and discussions}
In this paper we focused on the self-stabilizing token circulation and (local) leader election solutions in 
population protocol model augmented with agents and the eventual agent detector. 
The eventual agent detector 
eventually reports the presence or the absence of an agent.
%In the augmented model several solutions for leader election in rings and 
%complete networks have been proposed mainly in 
%\cite{DBLP:conf/opodis/FischerJ06, DBLP:conf/dcoss/AngluinACFJP05, 
%AngluinAFJ2005}. 
%In this work we extended the study to trees and arbitrary topologies.
We considered a very weak model of agents and nodes: anonymous, uniform and oblivious. 
Agents have no memory while nodes in the population 
have only one Boolean slot (not persistent).
In this model we proposed deterministic silent solutions for self-stabilizing 
local leader election. 
Moreover, we addressed the token circulation problem. 
Note that the agent paradigm materializes the token abstraction.
We proposed deterministic and probabilistic solutions and proved 
the necessity of the eventual agent detector even 
in environments helped by randomization.

The proposed model unifies several models for distributed 
interactions: population protocols, robots with global and 
local visibility (via the oracle paradigm) 
and the agents paradigm. Therefore, the current work opens several research directions. 
An interesting open issue would
be the study of the power of this model when agents and/or nodes have local memory.
Furthermore, an another interesting issue is to be explored the impact on the 
population stabilization of the full powered agents that execute 
some code when guested by a node.
\label{sec:conclusions}	
\singlespace
\bibliographystyle{IEEEtran}
\bibliography{Algo_Tree}

\section*{Annexes}
 
\subsection{Local leader election for even-sized rings}   
Intuitively the algorithm works as follows. 
A non-leader becomes a leader when  the responder is not a leader (Rule 2).
When two local leaders interact, the initiator becomes non-leader (Rule 1). 
%If the initiator is a leader and the responder is a non-leader, the latter becomes 
%leader and the former becomes non-leader (Rule 3). Otherwise, no state change occurs. 
%Each node outputs L when it holds a $\spadesuit$ , otherwise it outputs N.
Each node can hold either an agent, $\spadesuit$, or $-$.

\begin{algorithm}[h]
%\textbf {Algorithm 1}

\begin{verse}
\textit	{	\textbf	{Rule 1. } 	$((\spadesuit),( \spadesuit)) \longrightarrow  ((\spadesuit),(-)) $	}\\ 
\textit	{	\textbf	{Rule 2. } 	$((-),(-)) \longrightarrow  ((\spadesuit),(-)) $   }\\
%\textit	{	\textbf	{Rule 3. } 	$((\spadesuit,*),(-,*)) \longrightarrow  ((-),(\spadesuit))$ 	}
\end{verse}
\caption{Local leader election for even sized rings}
\label{alg:lle-even-rings}
\end{algorithm}

\begin{definition}
%{\bfseries Definition 1.} 
A configuration of Algorithm \ref{alg:lle-even-rings} is legitimate if in
each neighbourhood there is only one process holding an agent. 
This process will be called the local leader.
\end{definition}

\begin{lemma} Algorithm \ref{alg:lle-even-rings} is a silent implementation
of local leader election in even-sized rings using global fairness.
\end{lemma}

\begin{proof} 
Intuitively, the proof goes as follows.
Eventually, two neighbors interact and one of them becomes leader. 
Since the size of the ring is even the following cases can happen.
Either all nodes hold an agent or all of them are clean. In one round of 
interaction the system converges to a legitimate configuration (either applying Rule 1 or Rule 2).
All the other cases reduce to the above case.
\end{proof}

\end{document}